\documentclass[12pt]{article}

\newcommand{\bbr}{I\!\! R}
\newcommand{\bbz}{Z\!\!\! Z}

\newcommand{\3}{$^3$}

\newcommand{\x}{arXiv:}
\newcommand{\m}{\mathrm}

\usepackage{graphicx}

\begin{document}
\thispagestyle{empty}
\begin{center}

\null \vskip-1truecm \vskip2truecm {\Large{\bf

\textsf{Pre-Inflationary Spacetime in String Cosmology}

}}

\vskip1truecm {\large \textsf{Brett McInnes}} \vskip1truecm

 \textsf{\\  National
  University of Singapore}

email: matmcinn@nus.edu.sg\\

\end{center}
\vskip1truecm \centerline{\textsf{ABSTRACT}} \baselineskip=15pt
\medskip
Seiberg and Witten have shown that the non-perturbative stability
of string physics on conformally compactified spacetimes is
related to the behaviour of the areas and volumes of certain
branes as the brane is moved towards infinity. If, as is
particularly natural in quantum cosmology, the spatial sections of
an accelerating cosmological model are \emph{flat} and compact,
then the spacetime is on the brink of disaster: it turns out that
the version of inflationary spacetime geometry with toral spatial
sections is marginally stable in the Seiberg-Witten sense. The
question is whether the system remains stable \emph{before} and
\emph{after} Inflation, when the spacetime geometry is distorted
away from the inflationary form but still has flat spatial
sections. We show that it is indeed possible to avoid disaster,
but that requiring stability at all times imposes non-trivial
conditions on the spacetime geometry of the early Universe in
string cosmology. This in turn allows us to suggest a candidate
for the structure which, in the earliest Universe, forbids
cosmological singularities.

 \vskip3.5truecm
\begin{center}

\end{center}

\newpage

\addtocounter{section}{1}
\section* {\large{\textsf{1. Toral String Cosmology}}}

There are many reasons to suspect that, in string theory, the most
natural topology for the spatial sections of the Universe is that
of a \emph{flat, compact} three-manifold --- that is, the topology
of a torus or a certain finite quotient of a torus\footnote{If we
use a \emph{quotient} of a torus, the quotient should be
non-singular \cite{kn:reallyflat} for the ``large" dimensions,
though not necessarily for the ``small"
\cite{kn:greene}\cite{kn:lust}. Henceforth we refer to \emph{all}
such topologies as being ``toral".}. For example, consider the
recent work of Ooguri, Vafa, and Verlinde [\cite{kn:OVV}; see also
\cite{kn:dijk}], who have related the topological string partition
function to the Hartle-Hawking ``wave function of the Universe"
\cite{kn:hartle}. Ooguri et al begin by emphasising that it is
natural, in the context of quantum gravity, to assume that the
spatial sections of our universe are \emph{compact}, with their
initial size parameters being regarded as moduli to be selected by
a wave function with amplitudes peaked at the appropriate values.
Toral cosmology \cite{kn:zelda}\cite{kn:ruback} is particularly
natural in this picture, because the size of the torus is
decoupled from the spacetime length scale
--- it is not at all constrained by classical general relativity, and so it can \emph{only} be fixed by quantum gravity.

The idea of fixing the initial size of the Universe by means of the
wave function of the Universe has in fact been explored concretely
in \cite{kn:tye}\cite{kn:sarangi}\cite{kn:sashtye} [in the spherical
case] and \cite{kn:tallandthin} [in the toral case]; see also
\cite{kn:laura1} for related ideas. In particular, when the
Hartle-Hawking wave function [as modified by Firouzjahi, Sarangi,
and Tye \cite{kn:tye}\cite{kn:sarangi}\cite{kn:sashtye}] is applied
to the problem of predicting the initial size of a Universe created
from ``nothing", we find in the \emph{toral} case that the predicted
scale for the initial size of the Universe is about the string scale
\cite{kn:tallandthin}. This is a self-consistency check for our
claim that string theory favours toral sections: one does \emph{not}
reach this conclusion if locally spherical sections are assumed.

Toral cosmology also arises in the celebrated work of
Brandenberger and Vafa on \emph{string gas cosmology} [see
\cite{kn:brandvafa}, and \cite{kn:bat} for a recent extensive
review]. There the crucial arguments based on T-duality and string
winding modes make explicit use of the toral topology of
three-dimensional space. Still another string-cosmology
investigation in which toral sections play a basic role is the
recent work of McGreevy and Silverstein \cite{kn:silver1}.

Finally, of course, such spacetimes are also compatible with the
cosmological observations pointing to spatial flatness
\cite{kn:eisenstein}\cite{kn:tomo}, though, if there was a period
of Inflation, one does not expect to be able to observe the
non-trivial topology \emph{directly}. [It seems that this cannot
be done at the \emph{present} time \cite{kn:menzies}, and
inflationary theory suggests that it will probably never be
possible; see however \cite{kn:reb}.] However, the toral structure
may well have been ``observable" in the very early Universe, and,
as we shall see, this can yield crucial clues as to the spacetime
geometry at that time.

Our Universe appears to have a remarkable property: it passes
through phases in which its local geometry closely resembles that
of de Sitter spacetime. This appears to have happened in the past,
during an \emph{inflationary} era, and it seems likely that it
will happen again, due to the current presence of dark energy.

In toral cosmology, the inflationary era is described by
\emph{Spatially Toral de Sitter} [STdS] spacetime. If the torus
has side length 2$\pi$K at time t = 0, and if the spacetime
curvature is $-$1/L$^2$ [in signature ($+\,-\,-\,-$)],
corresponding to a positive vacuum energy density $\m{3/8\pi
L^2}$, then the metric is
\begin{equation}\label{eq:A}
g(\m{STdS})(\m{K,\,L)_{+---} \;=\; dt^2\;
-\;K^2\,e^{(2\,t/L)}\,[\,d\theta_1^2 \;+\; d\theta_2^2 \;+\;
d\theta_3^2]},
\end{equation}
where the torus is parametrized by angles; here t runs from
$-\,\infty$ to $+\,\infty$. This spacetime is \emph{locally}
identical to the familiar global de Sitter spacetime, but its
global structure is very different: the spacetime
topology\footnote{Note that K can be scaled to unity by a simple
translation of the time coordinate. This is however no longer true
if matter is inserted; see below for the physical significance of
K.} is $\bbr\,\times\,\m{T}^3$, whereas of course global simply
connected de Sitter spacetime has topology
$\bbr\,\times\,\m{S}^3$, where S\3 is the three-sphere. In this
work we shall explore the \emph{physical} differences that follow
from this \emph{topological} difference.

Physically, STdS spacetime differs from global [simply connected]
de Sitter spacetime in two fundamental ways: one pertaining to the
past, the other to the future.

First, as is well known, global de Sitter spacetime is
geodesically complete. \emph{This is not true of STdS spacetime}
\cite{kn:andergall}\cite{kn:gall}, which is incomplete in the
past. Furthermore, the incomplete region immediately develops into
a curvature singularity if any conventional matter is introduced;
and, worse yet, this behaviour is generic, in a strong sense that
we shall explain. But one of the main objectives of string
cosmology is to \emph{solve} the singularity problem in cosmology.
Clearly this contradiction needs to be addressed. It implies that
there must have been a \emph{pre-inflationary era} with a
spacetime geometry which must have involved significant
deformations of the STdS metric. In fact, the Andersson-Galloway
theorem, mentioned briefly in \cite{kn:tallandthin} and to be
discussed in detail here, tells us that the deformation must mimic
the effects of ``matter" which \emph{apparently} has negative
energy density.

The second difference is more obvious: the future conformal
boundary of global simply connected de Sitter spacetime is a copy
of S\3, while that of STdS is a copy of T\3. Now in string theory
it is known that the structure of conformal infinity has profound
consequences for the perturbative and especially the
non-perturbative stability of the theory. In particular, Seiberg
and Witten \cite{kn:seiberg} showed explicitly that having a
spherical structure at infinity plays a role in preventing
instability due to the nucleation of ``large branes" near
infinity. If one replaces the sphere at infinity by some other
space, then the theory is in danger of becoming inconsistent. This
Seiberg-Witten mechanism has been applied to topologically
non-trivial black hole spacetimes in \cite{kn:black}. More
importantly here, it has recently been applied to \emph{cosmology}
by Maldacena and Maoz \cite{kn:maoz}, and their work has been
extended in various ways in
\cite{kn:buchel}\cite{kn:porrati}\cite{kn:reallyflat}\cite{kn:unstable}.

The system remains stable near infinity as long as the conformal
structure at infinity continues to be represented by a metric of
positive \emph{scalar curvature}. If the scalar curvature at
infinity becomes negative, then the system is definitely unstable.
Thus the borderline between stable and unstable cases \emph{runs
somewhere through the space of conformally compactified manifolds
with zero scalar curvature at infinity}. It follows that replacing
global de Sitter spacetime with its toral version immediately
pushes the spacetime much closer to the brink of instability.

We have seen already that the spacetime geometry of the
pre-inflationary era must be a strongly deformed version of the
STdS metric, so that string cosmology can be non-singular.
Similarly, of course, the \emph{post}-inflationary era must have a
geometry which differs substantially from that of STdS, since
radiation and matter can no longer be ignored in that era. Given
that STdS is already perilously close to being unstable in the
Seiberg-Witten sense, \emph{it is clearly imperative to verify
that the pre- and post-inflationary distortions of the STdS metric
do not give rise to a drastic non-perturbative instability in
string theory}. This is one objective of the present work.

In order to explore this question, we need to specify more
precisely the ways in which the STdS geometry is deformed in the
pre- and post-inflationary eras. This is most straightforward in
the post-inflationary case, so we shall discuss that case first.
We find that the relevant brane action \emph{increases} towards
infinity after Inflation; hence it will always be positive,
indicating stability, provided that it was positive when Inflation
\emph{ended}. But conditions at the end of Inflation are
determined by conditions at its beginning.

The question of Seiberg-Witten stability at the beginning of
Inflation is much more difficult, because of uncertainties as to
the geometry at that time. Our strategy is to rely on the
constraints imposed by the fact that string cosmology should be
non-singular, and also by the very fact that \emph{it is not easy
to get Inflation to start at all} in string cosmology.

This point has recently been stressed by Linde, who argues
\cite{kn:lindetypical}\cite{kn:lindenew} that when Inflation is
embedded in string cosmology \cite{kn:racetrack}, it begins at a
fairly large length scale. But in string cosmology, the Universe
can be as small as the string length scale, which may be two
orders of magnitude smaller. Linde points out that discrepancies
like this could be a problem, because even if the Universe is born
in some very uniform state, it cannot in general be expected to
remain homogeneous during the time when it expands from the
string\footnote{Actually, Linde prefers a scenario in which the
Universe is born at the Planck scale.} to the inflationary length
scale. Linde proposes to solve this problem precisely by assuming
that the spatial sections of the Universe are toral. Under certain
circumstances, this allows all parts of the Universe to remain in
causal contact during the pre-Inflationary phase; then homogeneity
can be maintained by means of chaotic mixing \cite{kn:mixing}
until Inflation is ready to begin. Using this idea we can
construct a simple explicit model of the pre-inflationary era.

It turns out that the Seiberg-Witten brane action actually tends
to \emph{decrease} with the expansion during the pre-inflationary
era. Thus, even though its initial value is positive, there is a
danger that it will indeed become negative before the end of
Inflation. Because the inflationary era is long, measured in units
of the inflationary scale, even a gentle rate of decline can lead
to instability. This argument, combined with others, allows us to
put strong constraints on the spacetime geometry of the
pre-inflationary era, and on the nature of the [effective] field
with negative energy density. For example, we can rule out Casimir
energy as a means of resolving the initial singularity. The best
candidate appears to be the constraint field which appears in the
Gabadadze-Shang ``classically constrained gravity" theory
\cite{kn:gab1}\cite{kn:gab2}. This field is non-dynamical by its
very nature, and this naturally resolves the usual objections to
negative ``energies".

An alternative way of constraining the pre-inflationary geometry
was suggested in \cite{kn:tallandthin}. That approach is however
based on a specific modification of the Hartle-Hawking
wavefunction, the FST wave function \cite{kn:tye} [see also
\cite{kn:dealwis}\cite{kn:laura3}]. Here we take a much more
conservative approach based strictly on stability. It is pleasing
that, in fact, the results of the two approaches are consistent.

The principal new results of the present work are the following.
First, we show explicitly [Section 2] that there is no danger of
Seiberg-Witten instability in toral string cosmology as long as
conventional matter dominates. Next, while it is well known that
FRW models with \emph{exactly} flat spatial sections must
apparently violate energy conditions in order to be non-singular,
in Section 3 we use the results of Andersson and Galloway
\cite{kn:andergall} to argue that this conclusion is particularly
strong in the toral case: it remains valid even if we allow the
large distortions of spacetime geometry to be expected in the very
early Universe. Guided by this, in Section 4 we construct an
explicit one-parameter family of non-singular spacetimes which
allow us to investigate Seiberg-Witten instability in the
pre-inflationary era. Section 5 uses this explicit geometry to
identify the specific structure which allows the pre-inflationary
spacetime to be non-singular: the Gabadadze-Shang constraint
field.

\addtocounter{section}{1}
\section*{\large{\textsf{2. Stability After Inflation }}}

In this section we shall show that, even with toral spatial
sections, there is no danger of non-perturbative string
instability in the post-inflationary era, provided that there is
no such instability during and before Inflation. We begin by
briefly describing the toral versions of the relevant classical
cosmological spacetimes.

The post-inflationary Universe contains matter described by a
small positive cosmological constant [of, at least approximately,
the current value] together with various kinds of radiation and
matter. To discuss questions of stability it is simplest to follow
cosmological practice and regard this additional matter as a
general ``fluid" with an equation of state of the standard form
\begin{equation}\label{eq:JACKAL}
\m{p\;=\;w\,\rho},
\end{equation}
where p represents the pressure, $\rho$ the density, and where w
is a constant. It will be convenient to represent this constant in
terms of another constant $\epsilon$, defined by
\begin{equation}\label{eq:K}
\m{\epsilon\;=\;3\,(1\;+\;w)}.
\end{equation}
We emphasise that this ``fluid" is to be \emph{superimposed} on a
positive vacuum energy density of magnitude $\m{3/8\pi L^2}$: we
intend to introduce this substance into Spatially Toral de Sitter
spacetime.

Thus for example if we insert non-relativistic matter [zero
pressure] into the STdS spacetime and allow it to act [via the
Einstein equations], we shall have $\epsilon$ = 3, while
$\epsilon$ = 4 corresponds to ordinary radiation; values of
$\epsilon$ from 1 to 1.5 arise if we are interested in the
back-reaction induced by domain walls \cite{kn:domain} on the STdS
geometry, and so on. We stress however that we are \emph{not}
primarily interested in using this ``fluid" to violate the Strong
Energy Condition; in the cases of principal interest to us, the
acceleration is due to the negative contribution made by the STdS
cosmological constant to the total pressure. The ``fluid"
satisfies the Strong Energy Condition for all values of $\epsilon$
greater than or equal to 2.

We shall consider Friedmann cosmological models with metrics of
the form
\begin{equation}\label{eq:ALPHA}
g \;=\; \m{\m{dt}^2\; -\; \m{K^2\;a(t)}^2[d\theta_1^2 \;+\;
d\theta_2^2 \;+\; d\theta_3^2]};
\end{equation}
this generalizes the STdS metric in an obvious way. Adding the
energy density of the ``fluid" to that of the initial STdS space,
one can solve the Einstein equations to obtain
\begin{equation}\label{eq:EPSILON}
g\m{_s(\epsilon,\,K,\,L)_{+---} \;=\; dt^2\; -\;
K^2\;sinh^{(4/\epsilon)}\Big({{\epsilon\,t}\over{2L}}\Big)\,[d\theta_1^2
\;+\; d\theta_2^2 \;+\; d\theta_3^2]};
\end{equation}
here the s refers to the sinh function. The density of the
``fluid" turns out to be given by
\begin{equation}\label{eq:ZYNODIN}
\rho\;=\;{{3}\over{8\pi \m{L^2}}\,\m{a}^{\epsilon}}.
\end{equation}

If $\epsilon$ = 3, we should have the local metric for a spacetime
containing non-relativistic matter and a de Sitter vacuum energy,
and indeed $g\m{_s(\epsilon,\,K,\,L)_{+---}}$ reduces in this case
--- purely locally, of course --- to the classical Heckmann metric
[see \cite{kn:overduin} for a recent discussion]. In the general
case it agrees [again locally] with the results reported in
\cite{kn:eroshenko}. For large t we have
\begin{equation}\label{eq:THETA}
g\m{_s(\epsilon,\,2^{\,2/\epsilon}\,K,\,L)_{+---} \;\approx\;
dt^2\; -\;K^2\,e^{2\,t/L}[d\theta_1^2 \;+\; d\theta_2^2 \;+\;
d\theta_3^2]},
\end{equation}
which is the STdS metric given in equation (\ref{eq:A}); notice
that $\epsilon$ effectively drops out. Thus our metric is
``asymptotically STdS", for all $\epsilon$.

The matter content of these spacetimes does not behave as simply
as one might expect. For while it is true that both components,
the vacuum energy and the ``fluid", separately have constant
equation-of-state parameters, \emph{their combination does not}:
if we denote by W the ratio of the total pressure to the total
density, then we find
\begin{equation}\label{eq:IOTA}
\m{W\;=\;-\,1\;+\;{{\epsilon}\over{3}}\,
sech^2\Big({{\epsilon\,t}\over{2L}}\Big)}.
\end{equation}
Thus W decreases, from a limiting value of $-\,1\;+\;(\epsilon/3)$
as t tends to zero, to its asymptotic STdS value $-\,1$. The
Strong Energy Condition is violated if W $<$ $-\,1$/3, so [leaving
aside t = 0 itself, which we shall discuss shortly] the SEC holds
in the early universe provided that $-\,1\;+\;(\epsilon/3)\, >\,
-\,1/3$, which just means that $\epsilon$ should be greater than
2. In this case there is a transition from deceleration to
acceleration, as is observed in our Universe \cite{kn:riess}; so
this is the case of real physical interest.

This transition time occurs when t is precisely such that W =
$-\,1$/3, that is,
\begin{equation}\label{eq:IODINE}
\m{t_{TRANS}\;=\;{{2L}\over{\epsilon}}\,cosh^{-\,1}\Big(\sqrt{\epsilon/2}\Big)}.
\end{equation}
From equation (\ref{eq:EPSILON}) one can compute the circumference
C(t) of the spatial torus [along any component circle]; at the
transition time,
\begin{equation}\label{eq:ISLAND}
\m{C(t_{TRANS})\;=\;2\pi
K\,sinh^{(2/\epsilon)}\Big({{\epsilon\,t_{TRANS}}\over{2L}}\Big)\;=\;2\pi
K\,[(\epsilon/2)\;-\;1]^{1/\epsilon}}.
\end{equation}
This gives a physical interpretation of the length K: for example,
in the case of radiation [$\epsilon$ = 4], K is precisely the
radius of the spatial torus at the time of transition from
deceleration to acceleration. The point we wish to stress is that
K \emph{does} have a concrete physical meaning in these
spacetimes: it cannot simply be ``scaled away" by changing the
time coordinate, as can be done in pure STdS spacetime.

These spacetimes have a genuine [curvature] singularity at t = 0;
for example, the scalar curvature is given by
\begin{equation}\label{eq:IZOLA}
\m{R}(g\m{_s(\epsilon,\,K,\,L)_{+---}) \;=\;
-\,{{12}\over{L^2}}\;+\;{{3}\over{L^2}}\,(\epsilon\;-\;4)\,cosech^2\Big({{\epsilon
t}\over{2L}}\Big)};
\end{equation}
this tends to $-$12/L$^2$ as t tends to infinity, the correct
asymptotic de Sitter value in this signature, but it clearly
diverges as t tends to zero [except in the $\epsilon$ = 4 case,
which corresponds to radiation and hence to a traceless
stress-energy tensor which does not contribute to the scalar
curvature; but this case is still singular, as one sees by
examining other curvature invariants]. We stress that the
spacetime is singular for \emph{all} $\epsilon$, including values
such that the Strong Energy Condition is \emph{violated at all
times}. We shall discuss the significance of this basic fact in
detail in a later section.

Following Seiberg and Witten \cite{kn:seiberg}, we now turn to the
question of the \emph{non-perturbative} stability of these
spacetimes. The basic point here is that branes, being extended
objects, can be extremely sensitive to the geometry of the spaces
in which they propagate. It follows that modifying that geometry
can have major and unexpected consequences.

Seiberg and Witten consider the Euclidean version of the AdS/CFT
setup; this means that we are dealing with a space with a
well-defined conformal boundary, so that the geometry
asymptotically resembles that of hyperbolic space. Let us
introduce a 4-form field on a Euclidean, asymptotically hyperbolic
four-manifold, and consider the nucleation of Euclidean BPS
2-branes. The brane action consists of two terms: a positive one
contributed by the brane tension, but also a \emph{negative} one
induced by the coupling to the antisymmetric tensor field. As the
first term is proportional to the area of the brane, while the
second is proportional to the volume enclosed by it, we have in
four dimensions
\begin{equation}\label{eq:L}
\mathrm{S} \;=\;
\mathrm{T}(\mathrm{A}\;-\;{{\mathrm{3}}\over{\mathrm{L}}}\,\mathrm{V}),
\end{equation}
where T is the tension, A is the area, V the volume enclosed, and
L is the background asymptotic curvature radius.

The stability of this system is determined by a \emph{purely
geometric} question: can the area of a brane always grow quickly
enough to keep the action positive? If this is not the case, then
we have a serious instability, which has been described by
Maldacena and Maoz \cite{kn:maoz} as a \emph{pair-production
instability} for branes. In ordinary [non-compactified] hyperbolic
space the action is strictly positive, \emph{but in certain
distorted versions of hyperbolic space it is not}. Seiberg and
Witten found that this form of instability is never a problem near
infinity provided that hyperbolic space is deformed only mildly:
to be precise, the action remains positive near infinity as long
as the conformal structure at infinity continues to be represented
by a metric of positive scalar curvature [as is the case for
hyperbolic space itself, which has a \emph{spherical} conformal
boundary]. If the deformation is such that the scalar curvature at
infinity becomes negative, then the system is definitely unstable.
The borderline case, where the scalar curvature at infinity is
zero, \emph{is precisely the one of interest to us here}, since
the torus is flat, therefore scalar-flat. In this case, the
question of non-perturbative stability can only be settled by a
detailed examination; explicit examples of stable spaces with zero
scalar curvature at infinity have been given by Kleban et al
\cite{kn:porrati}, but explicit examples, with flat boundaries,
which are definitely \emph{unstable} have also been given in
\cite{kn:unstable}. The danger of instability is therefore very
real in this case.

Let us begin by examining the relevant Euclidean version of STdS
spacetime in detail. Because of the central role of the conformal
boundary in this discussion, it is natural to study the Euclidean
versions of our spacetimes by complexifying \emph{conformal} time
[as in \cite{kn:unstable} and the recent work of Banks et al
\cite{kn:banks}]. The Euclidean version of (\ref{eq:A}) is then
\begin{equation}\label{eq:B}
g(\m{STdS})(\m{K,\,L)_{++++} \;=\;
{{L^2}\over{\eta_+^2}}\,[\,d\eta_+^2\;+\;d\theta_1^2 \;+\;
d\theta_2^2 \;+\; d\theta_3^2]}.
\end{equation}
Here $\eta_+$, the Euclidean [dimensionless] conformal time, takes
its values in the open interval ($-\,\infty,\;0$). [K is now
hidden in the definition of $\eta_+$, which is given by
$\m{d\eta_+\;=\;dt/Ke^{t/L}}$.] Conformal infinity is defined by
extending to $\eta_+$ = 0, and clearly it is a copy of T$^3$ with
its standard conformal structure based on a \emph{flat} metric.
Thus spatially toral de Sitter spacetime lies close to the region
of Seiberg-Witten instability. The question is: how close?

One might think that, since the metric (\ref{eq:B}) differs from
that of hyperbolic space only globally, there should be no danger
of instability here. But branes, being extended objects, are
sensitive to certain global geometric features: the action
involves non-local quantities, namely area and volume, the growth
of which can be strongly affected by the global structure of the
ambient manifold. \emph{Changing the global structure might well
drive the system closer to the unstable case}. This does in fact
happen in the STdS case, though fortunately not to the extent that
any instability is actually induced.

We can compute the brane action [equation (\ref{eq:L})] for the
metric in equation (\ref{eq:B}) directly: the area form is just
$\m{-\,L^3\,d\theta_1\,d\theta_2\,d\theta_3/\eta_+^3}$ [since
$\eta_+$ is negative] and the volume form is
$\m{L^4\,d\eta_+\,d\theta_1\,d\theta_2\,d\theta_3/\eta_+^4}$. If
we imagine creating a brane at $\eta_+$ = E$_+ \,<\,0$ and then
moving it towards the boundary, the action as a function of
$\eta_+$ is
\begin{equation}\label{eq:M}
\mathrm{S_{STdS}(\eta_+) \;=\;
T\,\Big\{-{{8\,\pi^3\,L^3}\over{\eta_+^3}}\;-\;{{3}\over{L}}\,
\int_{E_+}^{\eta_+}{{8\,\pi^3\,L^4\,d\eta_+}\over{\eta_+^4}}\Big\}\;=\;-\,{{8\pi^3\,L^3\,T}\over{E_+^3}}\;>\;0}.
\end{equation}
Thus the action is \emph{a positive constant}. Notice that this
calculation only makes sense because we have compactified the
spatial sections; brane physics is indeed sensitive to the
distinction between ordinary de Sitter spacetime and its spatially
toral version.

Clearly the STdS spacetime is not unstable in the Seiberg-Witten
sense; however, it might not be difficult to render it unstable by
means of some arbitrarily small deformation, since even a small
negative contribution to the slope of the action could eventually
lead, as we move towards the boundary, to negative values for the
action itself. If it does become negative, then we have a
non-perturbative instability of the system. Explicit examples of
this have been given in \cite{kn:maoz} [for negatively curved
boundaries] and \cite{kn:unstable} [for flat boundaries].

Let us see how this works for the spacetimes we obtained earlier
by introducing various kinds of matter into STdS spacetime. As in
the case of STdS itself, we use conformal time. The Euclidean
version of our metric $g\m{_s(\epsilon,\,K,\,L)_{+---}}$ [equation
(\ref{eq:EPSILON})] that is asymptotic to
$g(\m{STdS})(\m{K,\,L)_{++++}}$ [equation (\ref{eq:B})] is
simply\footnote{Strictly speaking, K should be re-scaled by a
factor 2$^{(2/\epsilon)}$ for this statement to be precisely
correct.}
\begin{equation}\label{eq:N}
g\m{_s(\epsilon,\,K,\,L)_{++++} \;=\;
K^2\;sinh^{(4/\epsilon)}\Big({{\epsilon\,u(\eta_+)}\over{2\,L}}\Big)\,[\,d\eta^2_+\;+\;d\theta_1^2
\;+\; d\theta_2^2 \;+\; d\theta_3^2]}.
\end{equation}
Here the Euclidean analogue of conformal time, $\eta_+$, is
defined on the interval ($-\,$H(L/K,$\;\epsilon),\;$0), where, as
in equation (\ref{eq:B}), conformal infinity is at $\eta_+$ = 0,
and where
\begin{equation}\label{eq:IOTAIOTAIOTA}
\m{H(L/K,\;\epsilon)\;=\;{{2\,L}\over{K\epsilon}}\,\int_0^{\infty}\,{{dx}\over{sinh^{(2/\epsilon)}(x)}}}.
\end{equation}
This quantity is finite if and only if $\epsilon\,>\,2$; as we
saw, this is the observationally interesting range in which the
Universe undergoes a period of deceleration before beginning to
accelerate. Finally, u($\eta_+$) is the function of $\eta_+$ which
tends to zero as $\eta_+$ tends to $-\,$H(L/K,$\;\epsilon$) and
solves
\begin{equation}\label{eq:O}
\m{{{du}\over{d\eta_+}} \;=\;
K\,sinh^{(2/\epsilon)}\Big({{\epsilon\,u(\eta_+)}\over{2\,L}}\Big)}.
\end{equation}
The metric $g\m{_s(\epsilon,\,K,\,L)_{++++}}$ is an example of an
asymptotically hyperbolic Riemannian metric; we can think of it as
a deformation of the Euclidean STdS metric; as with the latter,
the underlying manifold is the product of an interval with T$^3$.
If we create a brane at $\eta_+$ = E$_+$ in the interval
($-\,$H(L/K,$\;\epsilon),\;$0), and consider the effects of moving
it towards infinity [that is, towards $\eta_+$ = 0], the relevant
action is
\begin{equation}\label{eq:P}
\m{S}(g\m{_s(\epsilon,\,K,\,L)_{++++}\,;\,\eta_+)\;=\;
8\pi^3\,TK^3\Big[\,sinh^{(6/\epsilon)}\Big({{\epsilon\,
u(\eta_+)}\over{2L}}\Big)\;-\;{{3}\over{L}}\int_{u(E_+)}^u
sinh^{(6/\epsilon)}\Big({{\epsilon \,u}\over{2L}}\Big)\,du \Big]}.
\end{equation}
In the case of non-relativistic matter [$\epsilon$ = 3] this can
be evaluated, the result being
\begin{eqnarray}\label{eq:Q}
\m{S}(g\m{_s(3,\,K,\,L)_{++++}\,;\,\eta_+)} = & \m{4\pi^3\,TK^3\Big[{{3u(\eta_+)}\over{L}}\;+\;e^{-\,3u(\eta_+)/L}\;-\;1} \nonumber\\
& +\;
\m{\;sinh\Big({{3u(E_+)}\over{L}}\Big)\;-\;{{3u(E_+)}\over{L}}\Big]}.
\end{eqnarray}
A sketch of the graph of the right side as a function of
u($\eta_+$) shows that this is \emph{never negative}, and so there
is no danger of Seiberg-Witten instability here.

In fact this is true for all values of $\epsilon$. A
straightforward calculation shows that the derivative of the brane
action in the general case is given by
\begin{equation}\label{eq:R}
\m{{{d}\over{d\eta_+}}\,S}(g\m{_s(\epsilon,\,K,\,L)_{++++}\,;\,\eta_+)\;=\;{{24\pi^3\,TK^4}\over{L}}
\,sinh^{(8/\epsilon)}\Big({{\epsilon\,
u(\eta_+)}\over{2L}}\Big)\Big[coth\Big({{\epsilon\,
u(\eta_+)}\over{2L}}\Big)\;-\;1\Big]}.
\end{equation}
This is obviously positive. Since
\begin{equation}\label{eq:S}
\m{S}(g\m{_s(\epsilon,\,K,\,L)_{++++}\,;\,E_+)\;=\;
8\pi^3\,TK^3\,sinh^{(6/\epsilon)}\Big({{\epsilon\,
u(E_+)}\over{2L}}\Big)}
\end{equation}
is also positive, it is clear that, \emph{provided we are in the
regime where this metric is valid}, the action will never be
negative --- there is no Seiberg-Witten instability here for any
value of $\epsilon$. However, we have to be cautious, for we know
that, in fact, the metric we are studying here is not valid in the
Inflationary era. Therefore, a more precise statement is as
follows: \emph{the action will never be negative if it is positive
when Inflation ends}, since it cannot decrease after that point.

Thus string cosmology \emph{is} stable, \emph{provided} that it is
stable up to the end of Inflation. The future of spacetime is
secure. The question is whether the same can be said of its
\emph{past}.

\addtocounter{section}{1}
\section*{\large{\textsf{3. The Past of Spatially Toral de Sitter and its Relatives}}}

In this section, we discuss the global theory of accelerating
spacetimes with toral spatial sections. It is well known that
\emph{exact} FRW cosmologies with \emph{exactly} flat spatial
sections must violate energy conditions if they are to be
non-singular. Less well known is the fact that these violations
are inevitable under \emph{very much more general} assumptions.

The problem of avoiding singularities in general accelerating
cosmologies is of course related to the question as to whether
Inflation must necessarily have a beginning. This has been
discussed extensively in the literature: see especially
\cite{kn:borde}. These discussions are based on FRW cosmological
models with spatial sections that are \emph{exactly flat} and
remain flat arbritrarily far back in time. In reality, of course,
the spacetime geometry could be severely distorted in a
pre-inflationary era, so the question of singularity avoidance
remained open. Recently, however, the work of Andersson and
Galloway \cite{kn:andergall} has given us a new insight into these
results, precisely in the case of interest to us here --- that of
compact spatial sections.

Consider again Spatially Toral de Sitter spacetime, with its
metric given in equation (\ref{eq:A}). Contrary to appearances,
this spacetime is actually geodesically incomplete: past-directed
timelike geodesics can wind around the torus in such a way that
they reach t = $-\,\infty$ in a \emph{finite} amount of proper
time. In this, of course, STdS differs radically from the version
of de Sitter spacetime with locally spherical spatial sections,
which is complete. The incompleteness here is relatively harmless,
in that it does not involve divergent curvatures; but we saw in
the previous section that the introduction of a wide variety of
kinds of matter into STdS spacetime \emph{inevitably} causes the
incompleteness to develop into a true curvature singularity. In
this, too, STdS differs from locally spherical de Sitter
spacetime, which remains non-singular even if [a small amount of]
conventional matter is introduced. Take, for example, the
spacetime corresponding to the Euclidean ``barrel" discussed in
\cite{kn:sarangi}, which is ``spatially spherical de Sitter plus a
small amount of radiation": it is non-singular.

We can summarize this state of affairs by saying that the toral
versions of de Sitter are ``more singularity-prone" than the
locally spherical versions.

To understand the reasons for this crucial property, which is of
basic importance for string cosmology, we must use the
Andersson-Galloway theorem \cite{kn:andergall}\cite{kn:gall}. The
theorem may be stated as follows; we refer the reader to the
Appendix for details of the terminology and a brief commentary on
this remarkable result.

\bigskip
\noindent THEOREM [Andersson-Galloway]: \emph{Let} M$_{\m{n+1}}$,
n $\leq$ 7, \emph{be a globally hyperbolic (n+1)-dimensional
spacetime with a regular future spacelike conformal boundary}
$\Gamma^+$. \emph{Suppose that the Null Ricci Condition is
satisfied and that} $\Gamma^+$ \emph{is compact and orientable.
If} M$_{\m{n+1}}$ \emph{is past null geodesically complete, then
the first homology group of} $\Gamma^+$, H$_1(\Gamma^+,\,\bbz)$,
\emph{is pure torsion}.
\bigskip

Here the \emph{Null Ricci Condition} is the requirement that, for
all null vectors k$^{\mu}$, the Ricci tensor should satisfy
\begin{equation}\label{eq:AARON}
\mathrm{R}_{\mu\nu}\,\mathrm{k}^\mu\,\mathrm{k}^\nu\;\geq\;0.
\end{equation}

The STdS spacetime has a regular future conformal boundary; this
boundary is compact and orientable, since it is just the torus
T$^3$. The Null Ricci Condition is satisfied, since the spacetime
is an Einstein space. But the first homology group of T$^3$ is
certainly \emph{not} pure torsion [that is, not every element is
of finite order]: it is isomorphic to
$\bbz\,\oplus\,\bbz\,\oplus\,\bbz$. Thus the spacetime \emph{had}
to be null geodesically incomplete. [In fact, as we mentioned
above, it is also timelike geodesically incomplete; see
\cite{kn:gall}.]

The surprising and beautiful feature of the Andersson-Galloway
theorem is that this same conclusion holds \emph{no matter how we
distort the geometry at early times}, as long as the Null Ricci
Condition continues to hold. That is, if we introduce \emph{any}
kind of matter at early times such that the NRC continues to hold
--- if the Einstein equations are valid, then this just means that
the Null Energy Condition [NEC] is satisfied --- then the
resulting cosmology will necessarily be null incomplete to the
past, no matter how distorted the spatial geometry may have been
at that time. In other words, the Andersson-Galloway theorem is a
\emph{singularity theorem}: as with the classical singularity
theorems, the conclusion does not depend on assumptions about
local isotropy or homogeneity or indeed on whether we assume a
Friedmann geometry at all. The prediction of a singularity is very
robust.

The Andersson-Galloway theorem is fundamental to string cosmology,
because one of the latter's major objectives is precisely to solve
the problem of the initial singularity in conventional cosmology.
The fact that the theorem makes such robust predictions means that
it tells us something very fundamental about these cosmologies.

What the Andersson-Galloway theorem is telling us --- see the
Appendix for a detailed discussion of this --- is that \emph{the
Null Ricci Condition [NRC] must be violated in the early history
of a non-singular accelerating string cosmology with toral
sections}.

Once we accept that NRC violation is inevitable here, we can use
this fact to construct an explicit spacetime metric for the
pre-inflationary era, and then use it to check for Seiberg-Witten
instability in that era.

\addtocounter{section}{1}
\section*{\large{\textsf{4. Stability Before Inflation: Not Guaranteed}}}
We have stressed that the \emph{only} way to solve the singularity
problem of accelerating string cosmologies is to violate the Null
Ricci Condition [NRC], or the Null Energy Condition [NEC] if we
assume the validity of the Einstein equations. We shall now use
this fact as a guide to the geometry of the earliest history of a
string cosmology: the pre-inflationary era.

A fundamental matter field which genuinely violates the NEC can be
hard to handle, for a variety of reasons ranging from simple
field-theoretic instability to conflict with the holographic
principle: see for example
\cite{kn:entropy}\cite{kn:carroll}\cite{kn:unstable}. Here we
shall take the conservative view that true NEC violation is not
acceptable in cosmology. Purported observations of violations of
the NEC at late times \cite{kn:NECVIOLATION} may have other
explanations; see for example
\cite{kn:csaki}\cite{kn:khoury}\cite{kn:kap}.

Instead of using fundamental matter fields violating the NEC, we
shall consider two alternative approaches. The first possibility
is that violations of the NRC in the early Universe are due to
energy densities which are \emph{negative}, not because of the
presence of some exotic matter field, but rather because unusual
topology or geometry or non-dynamical structures had equally
unusual effects at that time. A well-known example of such
phenomena is that the topology we are assuming here could lead to
Casimir effects; see for example \cite{kn:schaden}. Again, it is
widely agreed that negative-tension branes are acceptable if they
are not dynamical. More interestingly for our purposes here,
Gabadadze and Shang have pointed out that, in their ``classically
constrained" gravity theory \cite{kn:gab1}, the constraint field
can [in certain cases \cite{kn:gab2}] give rise to a negative
energy density. This is again physically acceptable since, by its
very definition, a constraint field will not be dynamical. We
shall return to this example below.

A second possibility [which can be combined with the first] is to
exploit the distinction between the NRC and the NEC. The former is
of course a purely geometric condition [see (\ref{eq:AARON})
above], while the latter refers only to items such as pressure and
energy densities. The two are linked by the gravitational field
equation. They are equivalent if the Einstein equations hold
exactly, but of course this is a highly questionable assumption in
the \emph{early} Universe. In braneworld models \cite{kn:varun}
there are explicit corrections to the Einstein equation which
allow the NRC to be violated \emph{while every matter field
satisfies the NEC}, so that one says that the NEC is
``effectively" violated \cite{kn:nojiri}\cite{kn:unstable}. The
distinction between the NEC and the NRC is also important when one
studies the relationship between brane and bulk when the brane
cosmology accelerates \cite{kn:apostol1}\cite{kn:apostol2}.
Similarly, the NEC and the NRC can be usefully different in
certain Gauss-Bonnet theories \cite{kn:sasaki}\cite{kn:gorby} and
also in scalar-tensor theories \cite{kn:polar}\cite{kn:uzan}; note
that scalar-tensor theories of precisely this kind do arise in
certain string cosmologies \cite{kn:kachru}. Notice that one can
think about the Gabadadze-Shang theory
\cite{kn:gab1}\cite{kn:gab2} in this way also, since this theory
certainly modifies the Einstein equation.

In short, violation of the Null \emph{Energy} Condition is not
necessarily the issue here. The real question is this: does
violating the Null \emph{Ricci} Condition have any direct
unwelcome consequences, even if the NEC is \emph{not} violated?

This is where Seiberg-Witten instability is central. For as we
have stressed, this particular form of instability depends only on
geometric data associated with branes [rates of growth of volume
and area]. Thus, unlike more familiar sources of instability,
Seiberg-Witten instability could be directly related to purely
geometric conditions like the NRC, and be entirely independent of
the gravitational field equations, whatever form the latter may
take for strong fields. That is, we might have to confront
Seiberg-Witten instability \emph{even under conditions where the
null energy condition is fully respected}. This in fact proves to
be the case.

In order to investigate the role of Seiberg-Witten instability in
the early Universe, we need an explicit metric describing the
geometry of the pre-inflationary era. We shall now construct a
family of [necessarily NRC-violating] relevant metrics, guided by
the special physics of string cosmology.

We shall assume that during the pre-inflationary era, as in the
inflationary era itself, there is a dominant form of matter with
positive energy density and which \emph{satisfies} the NEC; this
could be the inflaton itself. We assume that the energy density of
this component varies sufficiently slowly that it can be
approximated by the inflationary energy density
3/8$\pi$L$_{\m{inf}}^2$, where the inflationary length scale
L$_{\m{inf}}$ is of course very much smaller than the scale L we
used earlier to describe the late-time acceleration of the
Universe.

We shall assume that the NRC is approximately equivalent to the
NEC during the inflationary era, but that they differ
significantly in the pre-inflationary era. Thus, assuming again a
Friedmann metric of the form given in equation (\ref{eq:ALPHA})
above, during the inflationary era we have simply
\begin{equation}\label{eq:T}
\m{\Big({{\dot{a}\over{a}}\Big)^2 \;\approx
\;{{1}\over{L_{\m{inf}}^2}}}},
\end{equation}
whereas in the pre-inflationary phase we have
\begin{equation}\label{eq:U}
\m{\Big({{\dot{a}\over{a}}\Big)^2
\;=\;{{1}\over{L_{\m{inf}}^2}}\;+\;{{8\pi}\over{3}}\,\rho_{NRCViolating}\,}},
\end{equation}
where $\rho_{\m{NRCViolating}}$ simply measures the extent to
which the NRC fails; that is, it measures the disagreement between
the NRC and the NEC. Formally, we can think of it as an ``energy
density", but it does not necessarily correspond to any actual
matter field; nor will it obey the usual energy conditions.

Now string cosmology relies in a fundamental way on T-duality. For
example, in the specific case of the Brandenberger-Vafa theory,
the expansion of three dimensions away from the initial state is
explained by assuming a fluctuation in the form of an event
involving the annihilation of winding modes. One could object that
the \emph{creation} of winding modes, leading to
\emph{contraction}, is just as likely. But T-duality implies that
these must simply be two ways of describing the same event. This
means that a non-singular string cosmology is \emph{formally} a
``bounce" cosmology, but with a crucial difference: the
``contracting" phase of the history is just a redundant, dual
description of the expanding phase, not an actual physical era.

Thinking about string cosmology in this way makes it clear that
the initial state of the Universe must correspond to a spacelike
surface with \emph{zero extrinsic curvature}, like the surface of
maximal contraction in a true ``bounce" cosmology. We see at once
that this would not be possible if the strict Einstein equation
(\ref{eq:T}) held in the pre-inflationary era, but it is possible
with the modified equation (\ref{eq:U}), provided that
\begin{equation}\label{eq:V}
\m{{{8\pi}\over{3}}\,\rho_{NRCViolating}(0)\;=\;{{-1}\over{L_{\m{inf}}^2}}},
\end{equation}
where t = 0 at the initial state. That is, the initial total
density has to be zero, with the inflaton energy density being
exactly cancelled by the negative ``energy density" arising from
the disagreement between the NEC and the NRC, as discussed above.

Now we know that the NRC has to be violated here, and this means
that the total energy density must increase from its initial
value, zero. Since $\rho_{\m{NRCViolating}}$ is negative, this
just means that its absolute value must \emph{decrease} as the
Universe expands away from the initial state. A comparison with
the formula for the energy density in the post-inflationary era
[equation (\ref{eq:ZYNODIN})] suggests the ansatz
\begin{equation}\label{eq:W}
\rho_{\m{NRCViolating}}\;=\;{{-\,3}\over{8\pi
\m{L_{\m{inf}}^2}}\,\m{a}^{\gamma}}\,,
\end{equation}
where $\gamma$ is a positive constant and where the coefficients
have been chosen so that [equation (\ref{eq:V})] the scale factor
has unity as its initial value.

Obviously this way of representing the failure of the NRC in the
early Universe is far too simple to be realistic. Nevertheless we
shall see that it does capture the essential behaviour of the
Seiberg-Witten brane action in that era; furthermore it does
capture the behaviour of realistic candidates for the origin of
NRC violation. For example, if the NRC-violating component is a
simple Casimir energy arising from a toral topology
\cite{kn:coule}, we can use the form given in (\ref{eq:W}) with
$\gamma$ = 4. The parameter $\gamma$ \emph{is of basic importance
here}: it plays the same role in the pre-inflationary era that
$\epsilon$ [equation (\ref{eq:K})] plays in the post-inflationary
era --- that is, it fixes the ``equation of state". Our main
objective now is to understand how it is determined.

With any positive value of $\gamma$, the absolute value of the
NRC-violating ``energy density", $|\rho_{\m{NRCViolating}}|$,
automatically decreases with the expansion, and so the total
energy density [the sum of the positive vacuum energy and
$\rho_{\m{NRCViolating}}$] \emph{increases}: this means that the
NRC is violated overall. The ``pressure" corresponding to this
negative ``energy density" may be computed from the vanishing of
the covariant divergence of the total stress-energy-momentum
tensor, that is, from the equation
\begin{equation}\label{eq:X}
\m{{{d}\over{dt}}}\rho_{\m{NRCViolating}}\;+\;\m{{{3\,\dot{a}}\over{a}}}\,(\,\rho_{\m{NRCViolating}}\;+\;\m{p}_{\m{NRCViolating}})\;=\;0\,;
\end{equation}
the result is
\begin{equation}\label{eq:Y}
\m{p_{NRCViolating}\;=\;{{3}\over{8\pi
L_{inf}^2\,a^{\gamma}}}\,[\,1\;-\;{{1}\over{3}}\,\gamma]\,}.
\end{equation}
Combining this with equation (\ref{eq:W}), we see that the NRC is
indeed violated at all times in this geometry: this is consistent
with the Andersson-Galloway theorem.

Substituting (\ref{eq:W}) into (\ref{eq:U}) and solving, we obtain
\cite{kn:smash} a family of metrics parametrized by $\gamma$:
\begin{equation}\label{eq:PI}
g\m{_c(\gamma,\,K_0,\,L_{inf})_{+---} \;=\; dt^2\; -\;
K_0^2\;cosh^{(4/\gamma)}\Big({{\gamma\,t}\over{2L_{inf}}}\Big)\,[d\theta_1^2
\;+\; d\theta_2^2 \;+\; d\theta_3^2]},
\end{equation}
where the c subscript refers to the cosh function. Here K$_0$ is
simply the initial radius of the spatial torus [which is of course
very much smaller than the radius K discussed in previous
sections].

There are in fact two ways of thinking about this metric. Taking it
at face value, we have here a ``bounce" of the kind discussed in,
for example, \cite{kn:veneziano}. Such cosmologies have been
criticized [see for example \cite{kn:hertog}] on the grounds that
they demand extreme fine-tuning. An alternative approach is to
discard the contracting half of the spacetime, and investigate the
possibility of ``creation from nothing" \cite{kn:zelda} along t = 0.
If the initial torus is of the right size [about the string scale]
one might also interpret the remaining half of the geometry in terms
of string cosmology: by T-duality the ``contracting" phase of this
geometry merely represents a redundant description of the expanding
phase. In order to be definite we shall adopt a ``creation from
nothing at the string scale" interpretation here: so we are
\emph{only} concerned with values of t $\geq$ 0 in equation
(\ref{eq:PI}).

Since the function
$\m{cosh^{(4/\gamma)}\Big({{\gamma\,t}\over{2L_{inf}}}\Big)}$ is
[up to an overall constant factor] asymptotically independent of
$\gamma$ and indistinguishable from an exponential function, these
spacetimes are ``asymptotically STdS", with a toral conformal
boundary. Their \emph{past}, however, is very different from that
of STdS spacetime: as we hoped, these spacetimes are non-singular,
since the scale function never vanishes. We conclude that these
geometries are ideally suited to describe the pre-inflationary
era: they automatically generate an inflationary geometry in the
future, yet they are non-singular at the earliest times. The
Penrose diagram [Figure 1] is rectangular, with
width\footnote{Here $\chi$ represents any one of the angular
coordinates, which run from $-\,\pi$ to $+\,\pi$.} $\pi$ and
height
\begin{equation}\label{eq:MANGO}
\m{\Omega(L_{inf}/K_0,\,\gamma)\;=\;{{2\,L_{\m{inf}}}\over{\gamma\,K_0}}\,\int_0^{\infty}\,{{dx}\over{cosh^{(2/\gamma)}(x)}}}.
\end{equation}
The horizontal line at the bottom of Figure 1 represents the
creation of this universe [presumably from ``nothing", as in
\cite{kn:tye}\cite{kn:sarangi}\cite{kn:tallandthin}] at proper time
t = 0, which also corresponds to conformal time $\eta$ = 0. [That
is, this conformal diagram only represents the expanding half of the
geometry described by the metric in equation (\ref{eq:PI}).]
\begin{figure}[!h]
\centering
\includegraphics[width=0.7\textwidth]{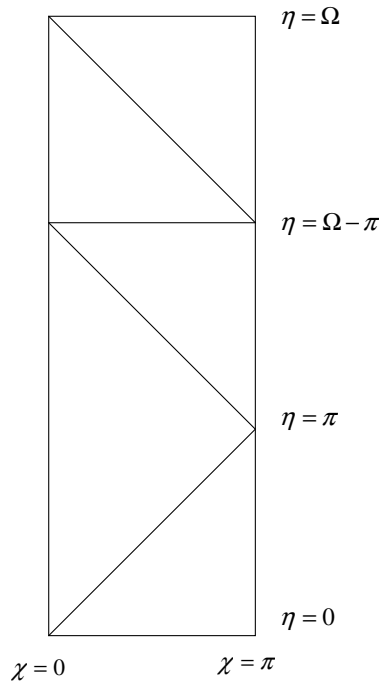}
\caption{Penrose diagram for the pre-inflationary/inflationary
eras.}
\end{figure}
Now we are interested in fairly \emph{large} values for the ratio
L$_{\m{inf}}$/K$_0$: we shall usually take it to be about 100. It
can be shown that, for all values of $\gamma$, this implies that
$\m{\Omega(L_{inf}/K_0,\,\gamma)}$ is similarly large. Thus,
bearing in mind that the Penrose diagram has a width equal to
$\pi$, we see that the diagram is very much taller than it is
wide. This means that our model gives a natural setting for
Linde's \cite{kn:lindetypical}\cite{kn:lindenew} solution of the
problem of getting low-scale Inflation \cite{kn:racetrack}
started. Linde observes that it is possible for the Universe to be
born with a significantly smaller size than the inflationary
scale, and yet still be sufficiently homogeneous for Inflation
actually to begin after the Universe expands to the appropriate
scale, provided that homogeneity can be maintained by means of
chaotic mixing \cite{kn:mixing} during the pre-inflationary era.
This idea was formulated in a more concrete way in
\cite{kn:tallandthin}, as follows.

Let us suppose that the initial spatial section was a torus of side
length 2$\pi$K$_0$, where the radius K$_0$ is the self-dual scale,
that is, the string length scale. The Universe expands in accordance
with the geometry given by the metric
$g\m{_c(\gamma,\,K_0,\,L_{inf})_{+---}}$, but global causal contact
is established provided that the Penrose diagram is \emph{tall
enough}. For, in that case, it will be possible for signals to
circumnavigate the entire Universe. In Figure 1, the lower rectangle
represents the pre-inflationary era subsequent to creation from
``nothing" at t = 0, while the upper square represents the era
during which the rate of expansion is so extreme that causal contact
begins to be lost: beyond the point $\eta$ = $\Omega \,-\,\pi$ in
the diagram, the most distant events can no longer affect local
physics. The conventional inflationary era must begin some time
after this point. As we have just seen, the values of
L$_{\m{inf}}$/K$_0$ in which we are interested here ensure that the
diagram should in fact be very much taller than it is wide, so there
is ample opportunity for a signal to be sent from a point on the
torus to the point most distant from it in this
direction\footnote{On a torus, the distance to the ``most distant
point" depends on direction. In practice our Penrose diagrams will
be so tall that this will not be an issue.} and back again. [Thus
the diagram in Figure 1 is as short as it can be; in reality, it
would be much taller.]

From Figure 1 we see that global causal contact begins to be lost,
in these directions, $\pi$ units of angular conformal time before
the top of the diagram [which represents the end of the era
described by this metric, that is, the end of Inflation]. In other
directions, global causal contact has already been lost by this
time. For Inflation to begin, the Universe must still be
sufficiently homogeneous, but we do not need it to be absolutely
homogeneous when Inflation begins. That is, we want a short gap
between the end of chaotic mixing and the start of inflation.

The geometry we are discussing here ensures this automatically, at
least for values of $\gamma$ that are not too small. To see this,
let T$_1$ be the proper time [in equation (\ref{eq:PI})] when
global causal contact begins to break down: this is given by
\begin{equation}\label{eq:FF}
\m{\pi\;=\;{{2\,L_{\m{inf}}}\over{\gamma\,K_0}}\,\int_{(\gamma
T_1/2L_{inf})}^{\infty}\,{{dx}\over{cosh^{(2/\gamma)}(x)}}}\,.
\end{equation}
Next, let T$_2$ be the proper time at which Inflation begins, that
is, the time at which the string scale [given by K$_0$] was
stretched to the inflationary scale [given by L$_{\m{inf}}$]. We
have then
\begin{equation}\label{eq:KYRAN}
\m{{{L_{\m{inf}}}\over{K_0}}\;=\;\m{cosh^{(2/\gamma)}\Big({{\gamma
T_2}\over{2L_{inf}}}\Big)}.}
\end{equation}
Judicious use of the elementary inequalities
$\m{{{1}\over{2}}e^x\,<\,cosh(x)\,\leq\,e^x}$, where x $\geq$ 0,
allows us to combine these relations to obtain
\begin{equation}\label{eq:GG}
\m{{{T_2}\over{L_{inf}}}\;>\;{{T_1}\over{L_{inf}}}\;+\;ln(\pi
/2^{(2/\gamma)})}.
\end{equation}
The second term on the right is positive provided that $\gamma$
exceeds about 1.21, as will be the case for all of the values we
shall discuss below. Thus Inflation does indeed begin after chaotic
mixing fails. The gap between the two events depends on $\gamma$. In
principle one could constrain $\gamma$ by imposing conditions on the
amount of inhomogeneity produced during the gap, but the present
model is too simplified for further investigation to be worth while
here; however, it might be of interest in more complex models.

Thus we have a simple model of the pre-inflationary era, one which
leads to a spacetime structure reflecting the basic physical
requirements: it is non-singular and is able to accommodate a
theory of initial conditions for Inflation. We claim that a
realistic theory of the earliest Universe would not lead to a
spacetime geometry vastly different from this one. It is therefore
appropriate to investigate the stability of this spacetime, in the
sense of Seiberg and Witten. The relevant family of Euclidean
metrics here is given by
\begin{equation}\label{eq:Z}
g\m{_c(\gamma,\,K_0,\,L_{inf})_{++++} \;=\;
K_0^2\;cosh^{(4/\gamma)}\Big({{\gamma\,u(\eta_+)}\over{2\,L_{inf}}}\Big)\,[\,d\eta^2_+\;+\;d\theta_1^2
\;+\; d\theta_2^2 \;+\; d\theta_3^2]},
\end{equation}
where as usual $\eta_+$ is Euclidean conformal time; we take it
that $\eta_+$ runs from $\m{-\,\Omega(L_{inf}/K_0,\,\gamma})$ [see
equation (\ref{eq:MANGO}); this corresponds to the smallest
transverse section, the Euclidean version of the initial spatial
section], to zero [which represents the conformal boundary, as
with the Euclidean conformal time coordinate of STdS itself ---
see equation (\ref{eq:B})]. The function u($\eta_+$) vanishes at
$\eta_+\,=\,\m{-\,\Omega(L_{inf}/K_0,\,\gamma})$ and satisfies
\begin{equation}\label{eq:AA}
\m{{{du}\over{d\eta_+}} \;=\;
K_0\,cosh^{(2/\gamma)}\Big({{\gamma\,u(\eta_+)}\over{2\,L_{inf}}}\Big)}.
\end{equation}

The Seiberg-Witten action for a brane of tension T created at the
smallest transverse section is
\begin{eqnarray}\label{eq:BB}
\m{S}(g\m{_c(\gamma,\,K_0,\,L_{inf})_{++++}\,;\,\eta_+)}\;= &
\;\m{
8\pi^3\,K_0^3\,T\Big[\displaystyle{\,cosh^{(6/\gamma)}\Big({{\gamma\,
u(\eta_+)}\over{2L_{inf}}}\Big)}}\; \nonumber\\
& -\;\m{\displaystyle{{{3}\over{L_{inf}}}}\displaystyle \int_0^u
cosh^{(6/\gamma)}\Big({{\gamma \,u}\over{2L_{inf}}}\Big)\,du
\Big]}.
\end{eqnarray}
Since u = 0 at the initial point, the action is initially
positive: the initial value is just $8\pi^3\mathrm{K_0^3\,T}$.
Unlike the action in the post-inflationary era [equation
(\ref{eq:P})], however, this action \emph{decreases} steadily as
we move towards infinity\footnote{Note that equation (\ref{eq:AA})
implies that
$\m{\lim_{\eta_+\,\rightarrow\,0}\,u(\eta_+)\,=\,\infty}$.}.
\emph{It is therefore far from clear that it will remain
positive.}

There are values of $\gamma$ which \emph{certainly do} lead to
Seiberg-Witten instability if the metric continues to have the
form (\ref{eq:PI}): for example, with $\gamma$ = 3 one has
\begin{equation}\label{eq:KERRY}
\m{S}(g\m{_c(3,\,K_0,\,L_{inf})_{++++}\,;\,\eta_+)\;=\;4\pi^3K_0^3\,T[1\;+\;e^{-\,3u(\eta_+)/L_{inf}}\;-\;{{3u(\eta_+)}\over{L_{inf}}}]},
\end{equation}
which becomes negative and is in fact unbounded below. However, we
shall see later that, for other values of $\gamma$, it is possible
for the action to decrease so slowly that it never becomes
negative. Thus the situation in the pre-inflationary era is much
more complex than in the post-inflationary era. In the latter
case, stability is guaranteed, for \emph{all} values of
$\epsilon$, provided of course that the brane action is positive
at the end of Inflation. Here, by contrast, we find that stability
is not guaranteed: it depends on the value of $\gamma$.

In fact, we can turn this argument around, and use the requirement
of stability to constrain $\gamma$. Once $\gamma$ is fixed, the
pre-inflationary geometry is likewise fixed. We now turn to this
question.

\addtocounter{section}{1}
\section*{\large{\textsf{5. Fixing the Pre-Inflationary Spacetime Geometry}}}
We saw in the previous section that certain values of $\gamma$
lead to a particular form of instability, while others do not. We
shall now argue that this is just one of several constraints on
$\gamma$; though in fact it will prove to be the strongest.

We begin with the following elementary observation. If we assemble
the ``energy density" $\rho_{\m{NRCViolating}}$ and the
``pressure" p$_{\m{NRCViolating}}$ into a four-vector, this vector
will be timelike or null precisely when $\gamma\,\leq\,6$, as can
be seen by comparing equations (\ref{eq:W}) and (\ref{eq:Y}).
Values of $\gamma$ greater than 6 correspond to an
``energy-momentum vector" which is \emph{spacelike}. Now we should
not immediately conclude that causality will be violated in this
case: a cosmological ``equation of state" is not a true equation
of state from which a ``speed of sound" can be deduced. [For
example, the speed of signal propagation in a quintessence field
is always locally equal to the speed of light, whatever the
``equation of state" may be \cite{kn:ratra}. See \cite{kn:odnoj}
for a general discussion of cosmological equations of state.] Even
granting this, however, and even granting that we do not
necessarily interpret $\rho_{\m{NRCViolating}}$ and
p$_{\m{NRCViolating}}$ in a literal way as energy density and
pressure, the presence of a spacelike vector in this context is
unwelcome\footnote{Note that this argument does not apply to the
\emph{total} energy density and pressure, because dark energy in
the form of a cosmological constant cannot ``carry a signal".};
one should view such an object with suspicion unless one has
physical arguments to the contrary [see for example
\cite{kn:turok}].

These suspicions are strongly supported by the fact that it can be
shown \cite{kn:tallandthin} that the action of the Euclidean
instanton defined by $g\m{_c(\gamma,\,K_0,\,L_{inf})_{+---}}$
diverges precisely when $\gamma\,>\,6$. Such values are therefore
ruled out by \emph{any} version of Euclidean quantum gravity. This
is the case, for example, whether one uses the Hartle-Hawking wave
function or some modification of it \cite{kn:tye}. We stress that,
because of this last remark, this argument is much more general
than the final conclusions reached in \cite{kn:tallandthin}, which
do depend on the details of the wave function.

In short, classical arguments and simple but general
quantum-gravitational considerations lead us to focus on values of
$\gamma$ which are less than or equal to 6.

In considering such values, we must bear in mind the fact that the
brane action given in (\ref{eq:BB}) is \emph{initially} positive;
it can only become negative at some point deeper into the space.
Now the metric (\ref{eq:PI}) is only supposed to describe the
geometry until the late inflationary era; for example, with a
conventional scalar field inflaton, when the kinetic term in the
inflaton lagrangian ceases to be negligible, and radiation begins
to become important, we must switch to a completely different
metric. That metric will in fact be approximated, at first, by the
metric $g\m{_s(4,\,K,\,L)_{+---}}$ [equation (\ref{eq:EPSILON})],
where K measures the size of the torus at the end of Inflation,
where L is the length scale determined by the \emph{current} value
of the cosmological constant, and where we have set $\epsilon$ = 4
to describe radiation. But we know that the brane action in the
Euclidean version of this geometry will actually \emph{increase}
[equation (\ref{eq:R})], and that this will continue to be the
case when matter begins to dominate over radiation [that is, when
$\epsilon$ = 4 is eventually replaced by $\epsilon$ = 3]. In other
words, the brane action will actually cease to decrease at some
point. It is conceivable that, if the action has been decreasing
sufficiently slowly up to that point, \emph{it may never have
become negative even if the expression given in (\ref{eq:BB})
indicates that it would ultimately do so if Inflation never
ended}. That is, what we should require is that the action should
remain positive up to this point, not necessarily everywhere.

Take for example the case of $\gamma$ = 3. Clearly [equation
(\ref{eq:KERRY})] the action in this case \emph{can} become
negative; in fact, a numerical investigation shows that it becomes
negative at a value of u($\eta_+$) given approximately by
0.4262$\,$L$_{\m{inf}}$. But a further numerical investigation
using equation (\ref{eq:FF}) shows that, assuming
$\m{L_{inf}\,\approx\,100\,K_0}$, the pre-Inflationary era ends at
a time corresponding to around u($\eta_+$) =
3.923$\,$L$_{\m{inf}}$ for $\gamma$ = 3. Thus, in this case, it is
clear that a severe instability will develop during the period
when the metric (\ref{eq:PI}) is still valid
--- that is, long before the end of Inflation. In fact, in this
particular case, the instability develops even before Inflation
begins. Thus $\gamma$ = 3 is certainly ruled out.

The case $\gamma$ = 4 is of particular interest, since that is the
value to be expected if the NRC-violating component is associated
with the Casimir effect \cite{kn:coule}. A numerical investigation
of (\ref{eq:BB}) in this case shows that, in this case, the action
becomes negative at u($\eta_+$) equal to about
0.4947$\,$L$_{\m{inf}}$, but (\ref{eq:FF}) implies, again assuming
$\m{L_{inf}\,\approx\,100\,K_0}$, that the pre-Inflationary era
ends at around 3.8070$\,$L$_{\m{inf}}$; thus, once again, the
system becomes unstable during the pre-inflationary era itself.
These conclusions do not change significantly if we take
$\m{L_{inf}\,\approx\,10\,K_0}$ --- the pre-inflationary era then
ends at about 1.5042$\,$L$_{\m{inf}}$, but this is still well
beyond the point where the action becomes negative. [If we take
$\m{L_{inf}\,\approx\,1000\,K_0}$, then the pre-inflationary era
ends at about 6.1096$\,$L$_{\m{inf}}$.] Thus it seems that Casimir
effects cannot account for the non-singularity of string
cosmology. This is an important and unexpected conclusion.
\begin{center}
\begin{tabular}{|c|c|c|c|}
  \hline
$\gamma$ & S = 0  & T$_1$ & T$_2$ \\
\hline
5.0000 & 0.6346  & 3.7377 & 4.8824\\
5.9000 &  1.2539 & 3.6954 & 4.8401\\
5.9900 &  1.9882 & 3.6919 &  4.8366\\
5.9990&  7.5654 & 3.6915 &  4.8363\\
6.0000&  $\infty$ & 3.6915 &  4.8362\\

  \hline

\end{tabular}
\end{center}

It turns out that the brane action vanishes at a point which
always recedes towards infinity as $\gamma$ increases. See the
Table; here T$_1$ and T$_2$ are as in equations (\ref{eq:FF}) and
(\ref{eq:KYRAN}), that is, they represent respectively the end of
the pre-inflationary era and the start of Inflation.
[$\m{L_{inf}\,\approx\,100\,K_0}$ is assumed throughout, and the
units are given by the inflationary length scale.] The rate at
which the zero point moves towards infinity is very slow at first;
the zero point is still within the region corresponding to the
pre-inflationary era even at $\gamma$ = 5.99. But the zero point
begins to move towards infinity at a rate which gathers pace
dramatically as $\gamma$ nears the value 6. Increasing $\gamma$
only slightly, from $\gamma$ = 5.99 to $\gamma$ = 5.999, causes
the zero point to jump well into the region corresponding to the
inflationary era. But this is of course still unacceptable: we
want the action to be positive throughout that region, that is,
for at least 60 e-folds from the point u($\eta_+$) = T$_2$. It is
clear that this can be done, but it is also clear that this will
require $\gamma$ to be extremely close to 6.

For $\gamma$ = 6 itself, the action can be evaluated explicitly:
it is given by
\begin{equation}\label{eq:EE}
\m{S}(g\m{_c(6,\,K_0,\,L_{inf})_{++++}\,;\,\eta_+)\;=\;8\pi^3K_0^3\,T\,e^{(-\,3u(\eta_+)/L_{inf})}},
\end{equation}
which is manifestly positive everywhere.

We saw earlier that there are strong physical arguments suggesting
that $\gamma$ can be no \emph{larger} than 6. Here we have seen
that brane physics demands that $\gamma$ can be only very slightly
\emph{smaller} than 6. Thus $\gamma$ is now tightly constrained.
We conclude that the spacetime geometry of the earliest Universe
can, within the simple model we are considering here, be
approximated by the metric
\begin{equation}\label{eq:DD}
g\m{_c(6,\,K_0,\,L_{inf})_{+---} \;=\; dt^2\; -\;
K_0^2\;cosh^{(2/3)}\Big({{3\,t}\over{L_{inf}}})\,[d\theta_1^2
\;+\; d\theta_2^2 \;+\; d\theta_3^2]},
\end{equation}
defined on a manifold of topology $\bbr\,\times\,$T\3. Here K$_0$
is the string scale and L$_{\m{inf}}$ is the inflationary scale.

The final picture is as follows. With $\gamma$ = 6, we have a
brane action which decreases from its initial positive value. The
decrease continues throughout the regions of the Euclidean space
corresponding to the pre-inflationary and inflationary eras. By
the point corresponding to end of Inflation, the action is
extremely small [equation (\ref{eq:EE})], \emph{but it is still
positive}. Beyond that point, the action begins to increase, and
it continues to increase [equation (\ref{eq:R})] indefinitely as
we move towards the boundary at infinity. Thus the action is
positive everywhere; there is no Seiberg-Witten instability in
this system.

In \cite{kn:tallandthin} we attempted to use the
Firouzjahi-Sarangi-Tye wave function
\cite{kn:tye}\cite{kn:sarangi} to predict the \emph{most probable}
value of $\gamma$, assuming that a spatially toral Universe could
be created from ``nothing" with any value of $\gamma$ greater
than\footnote{This restriction was based on the assumption,
confirmed here, that Inflation would never get started if $\gamma$
were smaller than or equal to 3.} 3. We found that, by an
overwhelming factor, the most probable value was precisely
$\gamma$ = 6. Thus, these two very different arguments agree. The
argument based on the FST wave function actually refines the
conclusion slightly, because it turns out that $\gamma$ = 6 is
very much more probable than any lower value, \emph{no matter how
close} it may be to 6: this happens because of a peculiar
discontinuity in the wave function at $\gamma$ = 6.

One important question remains: what is the physical nature of the
NRC-violating field we have been using here? We have seen that it
is \emph{not} a Casimir component, but what are the alternatives?
Inserting $\gamma$ = 6 into equations (\ref{eq:W}) and
(\ref{eq:Y}), we see that the pressure of the NRC-violating
component is negative and equal to the density; both density and
pressure decay rapidly with the expansion, according to the sixth
power of the scale factor. We should attempt to identify the kinds
of fields that can behave in this way.

One possibility was discussed recently by Patil \cite{kn:patil},
who points out that such densities and pressures can be produced
by a scalar field with a reversed kinetic term and a
\emph{vanishing} potential. The absence of any potential is
certainly very interesting in terms of string physics. However,
here we are not trying to use such densities and pressures to
replace the inflaton, only to supplement it in such a way that an
initial singularity is avoided. Furthermore the question of
stability is always difficult when this procedure is used.

An alternative that is more consistent with our approach here
appears in the work of Gabadadze and Shang \cite{kn:gab2}. As
mentioned earlier, the non-dynamical constraint field in
``classically constrained" gravity can lead to apparently negative
``energy densities" in a physically acceptable manner. In fact,
the Friedmann equation in this case \cite{kn:gab2} takes the form
[with flat sections]

\begin{equation}\label{eq:GABADAD}
\m{\Big({{\dot{a}\over{a}}\Big)^2
\;=\;{{1}\over{L_{\m{inf}}^2}}\;-\;{{b\,\varepsilon}\over{6\,a^6}}\,}},
\end{equation}
where b$\varepsilon$ is a certain constant which may be positive
or negative. In the positive case, with a suitable choice of
b$\varepsilon$, \emph{this is precisely} the equation (\ref{eq:U})
which we solved to obtain our metric (\ref{eq:DD}): the sixth
power is just what we need.

Thus we suggest tentatively that the NRC-violating structure in
the pre-inflationary era is just the Gabadadze-Shang constraint
field. The spatial sections are [compact] manifolds-with-boundary
in \cite{kn:gab2}, not tori; however, it should not be difficult
to reconcile these two views, since both spaces are flat and
compact. This appears to be a promising way of understanding the
physics of NRC violation --- and therefore of singularity
avoidance
--- in the early Universe.

\addtocounter{section}{1}
\section*{\large{\textsf{6. Conclusion }}}
The Andersson-Galloway theorems [see the Appendix] mean that it is
very difficult for an accelerating toral cosmology to be
non-singular. This is not a drawback: it merely implies that the
geometry of the earliest Universe must be strongly constrained.
Here we have tried to assess just how strong those constraints may
be in the context of a particular framework for string cosmology,
one based on Linde's suggestion that chaotic mixing may have a
crucial role to play in the earliest stages of Inflation.

What we find is that, with a few simple assumptions, the spacetime
geometry is completely fixed. The reader may wish to argue that
the model we have considered, based on the simple ansatz given in
equation (\ref{eq:W}), is in fact too simple --- though such
simplifications are standard practice in cosmology, and indeed
(\ref{eq:W}) is based directly on the familiar equation
(\ref{eq:ZYNODIN}). However, in \cite{kn:unstable} we studied the
more complex explicit NRC-violating spacetimes constructed by
Aref'eva et al \cite{kn:arafeva}; using the analysis developed
there, one can show that these spacetimes exhibit essentially the
same behaviour as the ones studied here. The general conclusion is
that a combination of very simple, \emph{essentially geometric}
requirements [formal energy-momentum vectors should not be
spacelike, actions of Euclidean instantons should not diverge, the
Seiberg-Witten brane action should not become negative] suffice to
impose surprisingly strong constraints on the pre-inflationary
spacetime geometry. The constraints are indeed strong enough to
allow us to specify the spacetime metric, and thus to suggest a
candidate for the origin of NRC violation: the Gabadadze-Shang
constraint field \cite{kn:gab2}.

It is striking that such strong conclusions flow from the basic
idea that the spatial sections of our Universe may have a
non-trivial topology. We have claimed that compact, flat manifolds
are best suited to quantum cosmology, but other compact candidates
\cite{kn:rp3}\cite{kn:silver2}\cite{kn:gab2}\cite{kn:psinas}
certainly merit further attention.

\addtocounter{section}{1}
\section*{\large{\textsf{ Acknowledgements}}}
The author is extremely grateful to Wanmei for preparing the
diagrams and for moral support. He wishes to thank Dr David Coule
for helpful correspondence, and he also sincerely thanks all those
who made possible his visit to the High Energy, Cosmology, and
Astroparticle Physics Section of the Abdus Salam International
Centre for Theoretical Physics, where the work described here was
initiated.

\addtocounter{section}{1}
\section*{\large{\textsf{ Appendix: About The Andersson-Galloway Theorem}}}
In this appendix we briefly explain the terminology used by
Andersson and Galloway \cite{kn:andergall}\cite{kn:gall}, and
comment on the conditions assumed in their theorem used above.
This is important, because our argument is based on the claim that
there is only one way to circumvent the theorem --- to violate the
Null Ricci Condition.

A four-dimensional spacetime M$_4$ with Lorentzian metric
$g_{\mathrm{M}}$ is said to have a {\em regular future spacelike
conformal boundary} if M$_4$ can be regarded as the interior of a
spacetime-with-boundary X$_4$, with a [non-degenerate] metric
$g_{\mathrm{X}}$ such that the boundary is {\em spacelike} and
lies to the future of all points in M$_4$, while $g_{\mathrm{X}}$
is conformal to $g_M$, that is, $g_{\mathrm{X}}$ =
$\Omega^2g_{\mathrm{M}}$, where $\Omega$ = 0 along the boundary
but d$\Omega \neq 0$ there. This is just a technical formulation
of the idea that the spacetime should be generic and de
Sitter-like at late times. There are examples of accelerating
spacetimes which do not have a regular future spacelike conformal
boundary, but these either severely violate the NRC at very
\emph{late} times \cite{kn:NECVIOLATION}, thus apparently
involving massive violations of the NEC over long periods of time,
or are highly non-generic, like Nariai spacetime \cite{kn:rp3}
[which is on the very brink of having a naked singularity]. One
certainly should not hope to escape from the conclusions of the
Andersson-Galloway theorem by resorting to such examples.

A spacetime is said to be \emph{globally hyperbolic} if it
possesses a Cauchy surface, that is, a surface on which data can
be prescribed which determine all physical fields at later and
earlier events in spacetime, since all inextensible timelike and
null curves intersect it. This forbids naked singularities, but it
also disallows ordinary AdS. The dependence of the
Andersson-Galloway theorem on this assumption might lead one to
ask whether our conclusions can be circumvented by dropping global
hyperbolicity. This is a topical suggestion, since it has recently
been claimed \cite{kn:veronika} that string theory allows
spacetimes which violate global hyperbolicity even more
drastically than AdS. To see why this too will not work here, we
need the following definition.

A spacetime with a regular future spacelike conformal completion
is said to be {\em future asymptotically simple} if every future
inextensible null geodesic has an endpoint on future conformal
infinity. This just means that there are no singularities to the
future --- obviously a reasonable condition to impose in our case,
since it would be bizarre to suppose that singularities to the
future can somehow allow us to avoid a Big Bang singularity.
Andersson and Galloway \cite{kn:andergall} show, however, that if
a spacetime has a regular future spacelike conformal completion
and is future asymptotically simple, \emph{then it has to be
globally hyperbolic}. Thus it would not be reasonable to drop this
condition in our context. Notice that this discussion has a more
general application: it means that, if the future of our Universe
resembles that of de Sitter spacetime, any attempt to violate
global hyperbolicity will necessarily cause a singularity to
develop to the future. It would be interesting to understand this
in the context of \cite{kn:veronika}.

Finally, one might wonder whether a compact flat three-manifold
can in fact have a first homology group which is pure torsion. The
rather surprising answer is that it can: there is a unique
manifold of this kind, the \emph{didicosm}, described in
\cite{kn:conway}\cite{kn:reallyflat}. The didicosm has the form
T$^3/[\bbz_2\;\times\;\bbz_2$], that is, it is a quotient of the
three-torus. However, if it were possible to construct a
singularity-free spacetime metric on
$\bbr\,\times\,\m{T}^3/[\bbz_2\;\times\;\bbz_2]$ without violating
the NRC, this metric would pull back, via an obvious extension of
the covering map T$^3 \rightarrow
\m{T}^3/[\bbz_2\;\times\;\bbz_2$], to a non-singular metric on
$\bbr\,\times\,\m{T}^3$, also satisfying the NRC. This is a
contradiction. Similarly, of course, one cannot escape the
conclusions of the theorem by allowing spacelike future infinity
to be non-orientable or by using orbifolds.

We conclude that the only physically reasonable way to avoid a
Bang singularity in an accelerating cosmology of toral spatial
topology is indeed to violate the NRC.

\end{document}